\documentclass[aps,prc,twocolumn,showpacs,superscriptaddress,floatfix]{revtex4}
\usepackage{graphicx}
\usepackage{longtable}
\usepackage{braket}
\usepackage{tabularx}
\usepackage{enumerate}
\usepackage{url}
\usepackage{enumitem}

\begin{document}
\title{Cross-shell excitations in doubly magic $^{132}$Sn and its nearest neighbours}
\author{Sangeeta Das}
\author {M. Saha Sarkar}\thanks{maitrayee.sahasarkar@saha.ac.in} \affiliation{Saha Institute of Nuclear Physics, HBNI, Kolkata-700 064, India}

\date{\today}

\begin{abstract}
Large scale shell model calculations have been performed to study the excitation spectra of $^{132}$Sn and its nearest neighbours   with a new 
cross-shell interaction constructed from two  widely used  interactions, $sn100pn$ and $CWG$,  of this mass region. A few of the two-body matrix 
elements have been  tuned to reproduce the low-lying multiplet states of  $^{132}$Sn. This is the first full scale shell model study of $^{132}$Sn energy 
spectra as well as transition probabilities. The excitation spectra for other nearest neighbours are reproduced reasonably well. The most important 
observable calculated are the E1 transition probabilities, which were so far beyond the scope of calculations with the existing interactions.
\end{abstract}


\maketitle

\section{\label{sec:level1}Introduction}
Nuclei near  doubly closed shell nuclei, act as testing ground of different nuclear models. Their excitation spectra \cite{nndc} typically show an admixture of two different modes of excitation. One mode arises from the excitation of valence particles within the shell and the other involves excitations across the shell gaps, which are especially important at higher excitation energies. 

In general, to describe the low energy states, different effective interactions are used, involving at best an intruder orbital from the higher opposite parity shell. The high energy states involving cross-shell excitations warrant the inclusion of other orbitals across the shell. However, such interactions involving cross-shell excitation possibilities are usually not available for heavy nuclei, apart from a few recent ones \cite{wang}. Moreover, unrestricted calculations over such large valence space encompassing two  major shells are mostly computationally challenging - thus, suitable truncation schemes are adopted.

The nuclei around the doubly magic $^{132}$Sn in A$\simeq$ 130 mass  region play a significant role in both nuclear structure physics and nuclear astrophysics. Thus this region has drawn the attention of experimentalists as well as theoreticians. Two different model spaces for neutrons are adopted to study the nuclei with neutron number $N$ $<$82 and $N>$82. The active neutron orbits in them are 
\begin {itemize}[leftmargin=*]
\item{} for $N$ $<$82, the neutron as well as proton model spaces consist of  the five orbits (1g$_{7/2}$, 2d$_{5/2}$, 2d$_{3/2}$, 3s$_{1/2}$ and  
1h$_{11/2}$) above $^{100}$Sn core with Z, N=50. 
\item{} for $N$ $>$82, the neutron model space includes six orbits, {\it viz.},  (1h$_{9/2}$, 2f$_{7/2}$, 3p$_{3/2}$, 3p$_{1/2}$, 2f$_{5/2}$, and 
1i$_{13/2}$), with proton model space same as above,  with Z=50 and N=82, i.e. $^{132}$Sn as the core.
\end {itemize}

The effective interactions $sn100pn$ and $CWG$, discussed in detail in Ref. \cite{brown1}, are obtained starting with a G matrix
derived from the CD-Bonn nucleon-nucleon interaction. They are widely used and  are successful in explaining the low energy states. However, they are unable to explain states at higher energies which involve cross-shell excitations. These interactions fail to explain the excitation spectra of a doubly closed shell nucleus, like, $^{132}$Sn and only its ground state can be calculated. Even for its very close neighbours, only a  few low energy states can be determined theoretically. These interactions are  quite successful in reproducing the excitation spectra, transition probabilities and moments of other nuclei. However, they fail to reproduce transition probability of hindered E1 transitions, because of the limitation in the valence space. There are no opposite parity orbitals in their model spaces with $\Delta I$= 1 to generate non-zero E1 transition density.
 
However, E1 transitions are frequently observed in these nuclei and play important roles in their low lying spectra. Not only in regions near the shell closures,  but also at the mid-shell (near $^{118}$Sn) region E1 transitions are observed as decay-out transitions of isomers \cite{nndc}. 

In the present work, we have derived a new effective interaction in an extended basis space including most of the neutron orbits from two major shells. The new interaction has been formulated to remove the limitations of the existing interactions.  The applicability of the interaction has been tested by calculating binding energies, excitation spectra and electromagnetic  transition probabilities of $^{132}$Sn and its close neighbours.

\section{\label{sec:level2} Derivation  of the effective interaction and Calculations}
In this section, we shall describe the derivation of the new interaction, which will be referred to as $sm56$ with appropriate suffixes for different variant interactions, like $sm56fp$ and $sm56fph$.

\subsection{\label{level2a} Model space and choice of single particle energies ($spe$)}
The model space consists of two proton ($\pi$) orbitals ($\pi$1g$_{7/2}$ and  $\pi$2d$_{5/2}$) and eight neutron ($\nu$) orbitals ( $\nu$1g$_{7/2}$ and  
$\nu$2d$_{5/2}$, $\nu$2d$_{3/2}$, $\nu$3s$_{1/2}$, $\nu$1h$_{11/2}$,  $\nu$1h$_{9/2}$, $\nu$2f$_{7/2}$, $\nu$3p$_{3/2}$) with $^{100}$Sn as core.

The present Hamiltonian has been constructed from two very well known interactions of this region: $sn100pn$ and $CWG$ interactions \cite{brown1}. For the sake of clarity in explaining this new interaction,  the proton orbits are denoted by P and the neutron orbits below N = 82 shell (1g$_{7/2}$, 
2d$_{5/2}$, 2d$_{3/2}$, 3s$_{1/2}$, 1h$_{11/2}$) are expressed as N1 and those above N = 82 shell 
(1h$_{9/2}$, 2f$_{7/2}$, 3p$_{3/2}$) are referred to as N2.
  
The proton single particle energies ($spe$)  are 0.8072 and   1.5623  MeV for $\pi$1g$_{7/2}$ and $\pi$2d$_{5/2}$  orbits.

The neutron single particle energies ($spe$)  are taken as -9.74, -8.97, -7.31, -7.62, -7.38 MeV for  the  orbits below N = 82 shell closure, {\it viz.}, 
$\nu$1g$_{7/2}$, $\nu$2d$_{5/2}$, $\nu$2d$_{3/2}$, $\nu$3s$_{1/2}$, $\nu$1h$_{11/2}$, respectively,  as suggested in Ref. \cite{brown1}. These $spe$s are chosen such that they reproduce the experimental levels of $^{131}$Sn  \cite{brown1}. 

For  the orbitals above N = 82 shell, we have  adjusted the neutron $spe$s. We have reproduced the experimental binding energy ($be$) of $^{133}$Sn with 
respect to $^{132}$Sn by adjusting the $spe$ of  $\nu$2f$_{7/2}$ orbit. All the orbitals below N = 82 shell are filled up and  only one particle is allowed to be in the orbit above N = 82 shell to reproduce the ground - state $be$, and  excitation energies of 3/2$^-_1$ and 9/2$^-_1$ states of 
$^{133}$Sn.  The $spe$s thus chosen  are 11.06, 5.718, 8.287 MeV for $\nu$1h$_{9/2}$, $\nu$2f$_{7/2}$,  $\nu$3p$_{3/2}$ orbits, respectively. 
  
\subsection{\label{level2b}The two body matrix elements ($tbme$s)}

The two-body matrix elements ($tbme$) are constructed in the following way.
\begin {itemize}[leftmargin=*]

 \item{} The proton-proton $tbme$s, $\braket{P_i P_j{\mid{V}\mid} P_k P_l}_{I,1}$ where i,j,k,l are  proton orbits $\pi$1g$_{7/2}$ and $\pi$2d$_{5/2}$ coupled to total angular momentum I and isospin T=1,  are taken from $sn100pp$ ($\pi-\pi$ part of $sn100pn$ interaction) including the Coulomb terms ($sn100co$). 
\item{} The neutron-neutron $tbme$s are been divided in a few groups. They are 
\begin {itemize}[leftmargin=*]
\item{}The $tbme$s of the intra-shell interaction for orbits below N = 82 shell, $\braket{N1{_i}N1{_j} {\mid{V}\mid} N1{_k}N1{_l}}_{I,1}$ 
(i.e for $\nu$1g$_{7/2}$, $\nu$2d$_{5/2}$, $\nu$2d$_{3/2}$, $\nu$3s$_{1/2}$, $\nu$1h$_{11/2}$) coupled to I and T=1, are obtained from $sn100nn$ interaction. 
\item{} For  N $>$ 82, $\braket{N2{_i}N2{_j} {\mid{V}\mid} N2{_k}N2{_l}}_{I,1}$  $tbme$s (i.e. interaction between $\nu$1h$_{9/2}$, $\nu$2f$_{7/2}$,  $\nu$3p$_{3/2}$) coupled to I and T=1, have been taken from $CWG$ interaction. 
\item{} The T = 1, $\pi-\nu$ $tbmes$ of $CWG$ interaction have been taken as cross-shell neutron-neutron $tbme$s, {\it viz.,} $\braket{N1{_i}N2{_j} {\mid{V}\mid} N1{_k}N2{_l}}_{I,1}$ keeping in mind  the charge independence and isospin invariance of nuclear interaction.
\item{} The rest of neutron-neutron $tbme$s like $\braket{N1{_i}N2{_j} {\mid{V}\mid} N1{_k}N1{_l}}_{I,1}$ or $\braket{N1{_i}N2{_j} {\mid{V}\mid} N2{_k}N2{_l}}_{I,1}$, which are not present in the $\pi-\nu$ $tbme$ set of $CWG$ interaction have been calculated from zero range delta interaction \cite{heyde}. 
\end{itemize}
\item{} The proton-neutron $tbme$s for neutron orbits below N=82, {\it viz.}, $\braket{P_iN1{_j} {\mid{V}\mid} P_kN1{_l}}_{I,1 ~{\rm and} ~0}$ have been considered from $sn100pn$ and those for $\nu$ orbits above N = 82 shell,  {\it viz.}, $\braket{P_iN2{_j} {\mid{V}\mid} P_kN2{_l}}_{I,1 ~{\rm and} ~0}$ have been taken from $CWG$ interaction.  
\item{} Rest of the $\pi-\nu$ $tbme$s,  {\it viz.}, $\braket{P_iN1{_j} {\mid{V}\mid} P_kN2{_l}}_{I,1 ~{\rm and} ~0}$ have been calculated from delta interaction as discussed above.
\end{itemize}

\subsubsection{\label{level2aa}Tuning of cross-shell $\nu-\nu$ $tbme$s}
However, the set of cross-shell $\nu  - \nu$ $tbme$s of particle-hole multiplets, which are obtained from the T=1 $tbme$s of $\nu-\pi$ interaction from $CWG$ Hamiltonian can not reproduce the low-lying experimental spectra of $^{132}$Sn \cite{131sn_p_bhatta_132sn}. These states  are identified by earlier workers \cite{133sb_bocchi_2016} as members of particle-hole multiplets from theoretical calculations using Random Phase Approximation (RPA).

This failure of the collated interaction is not unexpected. Although the nuclear interaction is charge independent - the requirement of Pauli exclusion  principle which warrants anti-symmetrization of the two-nucleon wave function, distinguishes the interaction between like nucleons from that arising
 between  unlike nucleons. It is found that although the interaction energies of T=1 parts of $\nu-\nu$ , $\nu-\pi$ and $\pi-\pi$ are alike for a pair of 
nucleons in the same orbit, the monopole interaction between like nucleons is weaker than those between unlike ones \cite{sorlin}. For dissimilar orbits, 
the interaction energies are grossly different.  For two unlike nucleons, the monopole interaction is attractive. However, for like nucleons, it is mainly repulsive.

Thus to tune the $\nu  - \nu$ $tbme$s of important and relevant multiplets for $^{132}$Sn, the procedure described in Ref. \cite{sorlin} has been followed. The experimental binding energies of $^{131}$Sn, $^{132}$Sn and $^{133}$Sn from  Ref.  \cite{ame} and the experimental excitation energies of multiplet states  are utilized to get the diagonal $tbme$s of corresponding multiplets.

\begin{figure}
\includegraphics[width =1\linewidth]{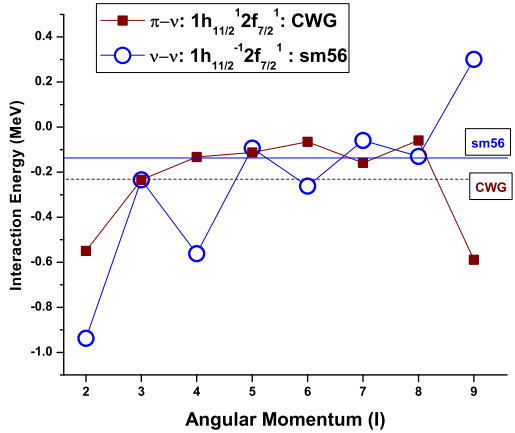}

\caption{\label{sm_2}Empirical interaction energy between two nucleons. One of them is in 1h$_{11/2}$ and other in 2f$_{7/2}$ orbit. The $CWG$ ($sm56$) $tbme$s are for $\pi-\nu$ 
($\nu-\nu$) interaction.  The dashed (solid) line indicates the monopole energy for $CWG$ ($sm56$) interaction.}
\end{figure}

\begin{itemize}[leftmargin=*]
\item{\it The low-lying positive parity states in $^{132}$Sn:}

The two cross-shell positive parity multiplets which are of utmost importance in the low-lying spectra of $^{132}$Sn arise from couplings of 
 1h$_{11/2}^{-1}$ -2f$_{7/2}$  \cite{133sb_bocchi_2016} and 1h$_{11/2}^{-1}$ -3p$_{3/2}$ orbitals.  The coupling of 1h$_{11/2}^{-1}$ -1h$_{9/2}$ orbits has been the most significant  for the generation of $8^+_2$ state.

\begin{itemize}[leftmargin=*]
\item {\it The $\nu 1h_{11/2}^{-1}$ -$\nu 2f_{7/2}$ $tbme$s :}\\
The experimental binding energies of $^{131}$Sn, $^{132}$Sn and $^{133}$Sn from  Ref.  \cite{ame} and the experimental excitation energies of $^{132}$Sn (2$^+_1$, 4$^+_1$, 5$^+_1$, 6$^+_1$, 7$^+_1$, 8$^+_1$, 9$^+_1$, which are considered to be members of 1h$_{11/2}^{-1}$ -2f$_{7/2}$ multiplet \cite{133sb_bocchi_2016}) are utilized to get the diagonal  $tbme$s of $1h_{11/2}^{-1} - 2f_{7/2}$ coupling. Since the minimum angular momentum generated by this coupling is 2, the E$_x$(2$^+$)= 4.041 MeV is used as reference energy. Most of the $tbme$s are negative and thus the interaction is attractive 
(Fig.\ref{sm_2}).  The   monopole contribution to the interaction energies  is denoted by the dashed line in the figure. The adopted experimental level scheme of $^{132}$Sn does not include a 3$^+$ state, thus the corresponding interaction energy has not been modified. The Fig.  \ref{sm_2} shows  comparison of the $\pi-\nu$ interaction energies from  $CWG$ interaction  \cite{brown1} with the present modified ones for $\nu-\nu$. Similar to the observation in Ref. \cite{sorlin}, for  $1h_{11/2}^{-1} - 2f_{7/2}$ coupling, the monopole interaction between neutrons (-0.1374 MeV) is about two times less attractive than those between neutron-protons (-0.2317 MeV). 

  \item {\it Tuning of $\nu1h_{11/2}^{-1}$ -$\nu3p_{3/2}$: }\\
The $\nu1h_{11/2}^{-1}$ -$\nu3p_{3/2}$ coupling can generate angular momenta $4^+$ to $7^+$. At the low spins, experimental level scheme of $^{132}$Sn,
only two $6^+$ and $7^+$ are reported. Unlike those for $\nu 1h_{11/2}^{-1}$ - $\nu 2f_{7/2}$ coupling, all $tbme$s corresponding to $\nu1h_{11/2}^{-1}$ coupled with  $\nu3p_{3/2}$   could not be determined systematically, due to lack of experimental data. However, only two relevant $tbme$s are tuned to improve the predictions for second $6^+$ and $7^+$ states. 

\item {\it Tuning of $\nu1h_{11/2}^{-1}$ -$\nu1h_{9/2}$  $tbme$s: }\\
Similar situation prevails for tuning the $tbme$s of the coupling of $\nu1h_{11/2}^{-1}$ -$\nu1h_{9/2}$, which  can generate angular momenta
 $1^+$ to $10^+$. Relevant two $tbme$s for this coupling have been tuned to  improve the predictions for second $7^+$ and $8^+$ states. 

\end{itemize}

\item{\it The low-lying negative parity states in $^{132}$Sn:}

The negative parity  $3^-_1$, $4^-_1$ and $5^-_1$ states do not have  pure multiplet structure. These angular momenta can be generated from the coupling of $\nu2d_{3/2}^{-1}$ and $\nu 2f_{7/2}$. The $3^-_1$, $4^-_1$ states also have contributions from the $\nu3s_{1/2}^{-1}$  coupled with $\nu 2f_{7/2}$. The holes in $\nu2d_{5/2}$ and $\nu1g_{7/2}$ can couple with $\nu 2f_{7/2}$, to generate these negative parity states. Thus these  matrix elements are 
not modified. 
 \end{itemize} 

\subsubsection{\label{level2ab}Tuning of inter-shell  $tbme$s}
The new modified interaction does not include the full proton model space as considered in $sn100pp$ and $CWG$ interactions. The neutron model space 
for N$>$82 has also been truncated from the one used in $CWG$ Hamiltonian. Thus the corresponding $\pi-\pi$ and $\nu-\nu$ $tbme$s of the truncated 
spaces are modified.
\begin{itemize}[leftmargin=*]

\item{\it The $\pi-\pi$ tbmes: }\\ The diagonal two body matrix element for $\pi 1g_{7/2}$-$\pi 1g_{7/2}$ coupled to I=0 is  adjusted to reproduce the binding energy  of $^{134}$Te.

\item{\it The $\nu-\nu$ tbmes: }\\ The $\nu 2f_{7/2}$- $\nu 2f_{7/2}$ $tbme$s are adjusted to reproduce the binding energy and excitation energies 
of $2^+_1$ to $6^+_1$ levels of  $^{134}$Sn. The $\nu 2f_{7/2}$- $\nu 1h_{9/2}$ $tbme$ corresponding to I=$8^+$ has been adjusted to reproduce the  experimental $8^+_1$ excitation energy.
\end{itemize}

\subsection{\label{level2b}The interactions}
Two versions of the interaction have been generated. The first one, named as $sm56fp$ includes only the modifications in  $\nu-\nu$ $tbme$s 
(except those corresponding to $\nu 1h_{11/2}^{-1}$ -$\nu 1h_{9/2}$) of the collated interaction. 
 The final version  of the interaction which includes all modifications in $\nu-\nu$ and $\pi-\pi$ $tbme$s as discussed above, is named as $sm56fph$.

\subsection{\label{level2b} The calculations}

\begin{figure}
\includegraphics[width =1.5\linewidth]{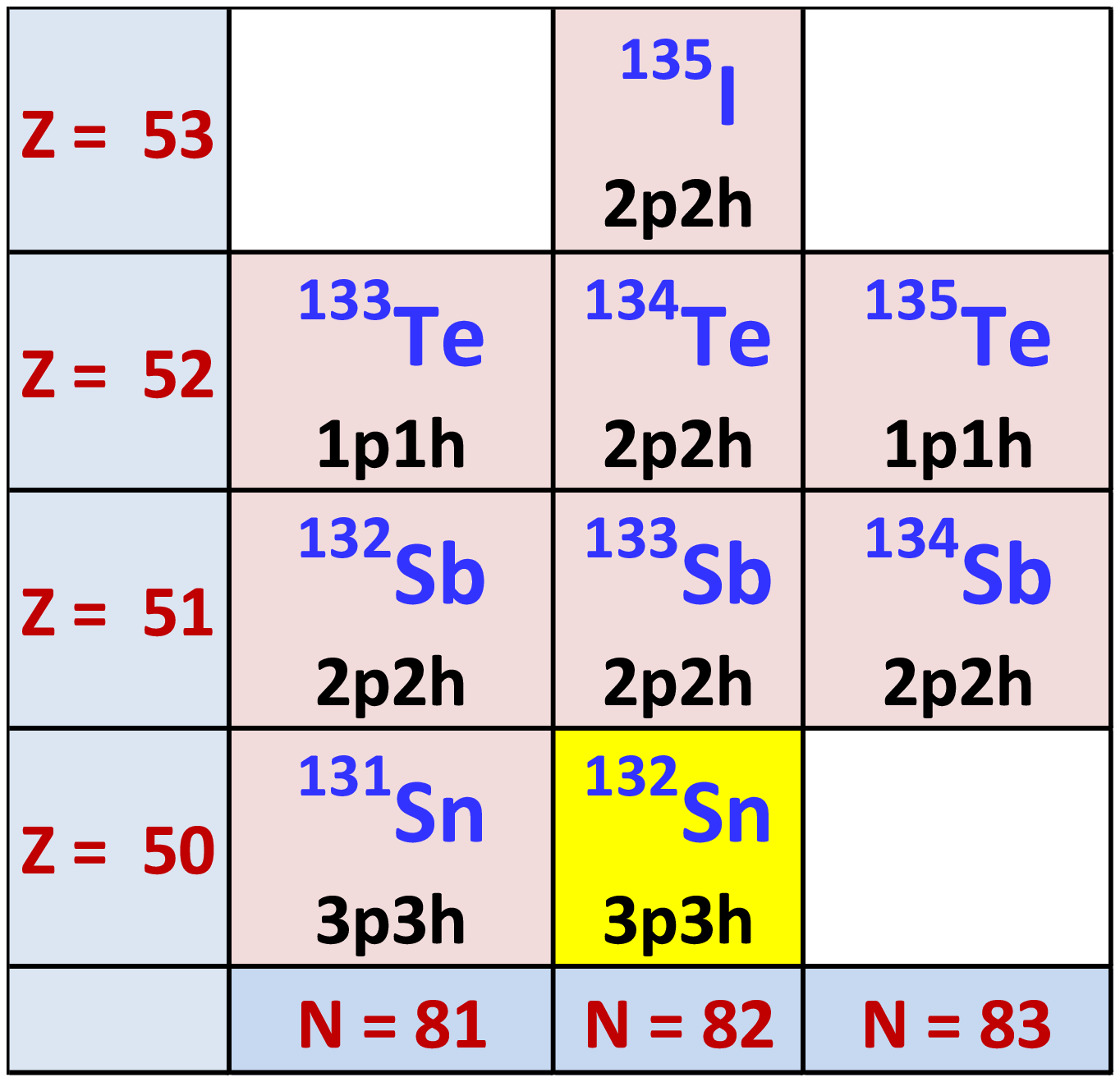}
\vspace{-3cm}
\caption{\label{nuclei}List of nuclei for which theoretical calculations have been done. The maximum npnh excitation mode which was computationally possible in the presesnt calculation has been indicated in the chart.}
\end{figure}

 The shell-model code NUSHELLX@MSU \cite{nushellx} and OXBASH \cite{brown2} are used for the calculations using the the interactions discussed above. 
The   excitation spectra as well as transition probabilities (including B(E1) values) of $^{132}$Sn have been calculated and compared with experimental 
data. The predictability  of our shell-model Hamiltonian in the neighbourhood of $^{132}$Sn has also been tested. The excitation spectra of  a few nuclei 
around $^{132}$Sn with neutron numbers 81 - 83 are calculated. They are $^{131,132}_{~~~~~50}Sn_{81,84}$, $^{132-134}_{~~~~~~~51}Sb_{81-83}$,  
$^{133-135}_{~~~~~~~52}Te_{81-83}$ and $^{135}_{~53}I_{82}$.  The excitation spectra of these nuclei have been calculated with  no restriction of neutrons for N $<$ 82 orbitals. For nuclei with N$\leq$82 neutrons, allowed excitation modes considered are: (i)  no neutron (0p0h) beyond N$>$82 orbits,
 or, (ii) at least one neutron excited in any of the three orbits (1p1h), (ii) 2p2h and (iii)  maximum one neutron excited in each of the three orbits (3p3h). Thus from no neutron in N$>$82 orbits, 1p1h to 3p3h excitations in these orbits are possible in this choice. For nuclei with N$>$82, 
apart from the neutrons already in N$>$82 orbitals, 1p1h to 3p3h excitations are allowed. However, in some of the cases discussed below, such options were not computationally  feasible with our facilities. Thus we had to restrict to 2p2h or even 1p1h excitations only.  A  chart (Fig. \ref{nuclei}) shows the  nuclei for which theoretical calculations have been done. The maximum $npnh$ excitation mode which was computationally possible in the present calculation is also indicated for each nucleus in the chart.

\section{\label{sec:level3}Results and Discussion}

\subsection {\label{level3a}Binding Energies}
\begin{figure}
\includegraphics[width =1.\linewidth]{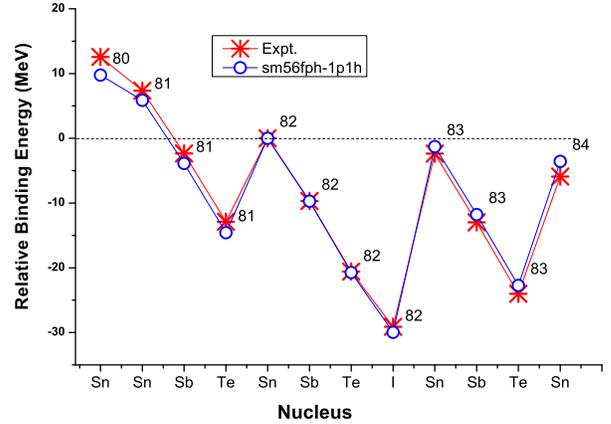}

\caption{\label{mass} Comparison of calculated and experimental binding energies with respect to $^{132}$Sn.}
\end{figure}

The trend in the calculated binding energies in 1p1h option for neutrons (where maximum one neutron can be excited to any one of the orbits above 
N=82 and no restriction in the orbits below N=82) with respect to  that of $^{132}$Sn are reproduced well in theory (Fig. \ref{mass}). The values are plotted for isotones of $Sn$ to $I$ for N=80 to 84. The excitation spectra of these nuclei have been calculated with the  $sm56fph$ interaction. 
 
\subsection {\label{level3b}Excitation Energies}
\begin{itemize}[leftmargin=*]
\item {\it Calculations for N=82 isotones } 
\subsubsection{\label{level3ba}$^{132}_{50}Sn_{82}$} 
The excitation spectra of $^{132}$Sn has been calculated with  no restriction of neutrons for N $<$ 82 orbitals with at best one 
particle excited in each of the three orbits for N $>$ 82. Thus from no neutron in N$>$82 orbits, 1p1h to 3p3h excitations in these orbits are
allowed.
 
The doubly closed shell $^{132}$Sn with Z=50 and N=82, has its  first excited state, 2$^+$ at relatively higher energy($\sim$ 4 MeV) than its neighbouring nuclei due the excitation of neutrons across the N=82 shell closure. Thus, the low lying excited states of this doubly closed shell nucleus can be explained by particle-hole (p-h) excitations. As the excitations  involve  only a few particle and hole states, the structure of individual states should possess nearly pure multiplet structure. 

P. Bhattacharyya {\it et al.} \cite{131sn_p_bhatta_132sn} have identified  the first 2$^+$, 4$^+$, 5$^+$, 6$^+$, 7$^+$, 8$^+$, 9$^+$ states  as members of $\nu$1h${_{11/2}}{^{-1}}$2f${_{7/2}}{}$ multiplet. In the present work, the interaction has been tuned as discussed in Sec. \ref{sec:level2}. The theoretical results from 3p3h excitation modes with $sm56fp$ and $sm56fph$ interactions show similar agreement  with  experimental data in 
Fig. \ref{132sn}, except for the 8${^+}_2$ state.  

 Fogelberg {\it et al.} \cite{Fogelberg_132sn} have described the positive parity  6${^+}_2$, 8${^+}_2$,  and 7${^+}_2$ states near 5.4 MeV 
as members of  $\pi$1g${_{9/2}}{^{-1}}$1g${_{7/2}}{}$ multiplet. The $\pi1g{_{9/2}}$ orbital is not included in our valence space. 
However, in the present calculation, 6$^+_2$, and 7${^+}_2$ are reproduced well with $sm56fp$ and they originate from $\nu1h{_{11/2}}{^{-1}}2p{_{3/2}}{}$ multiplet. The 8${^+}_2$ corresponding  to $\nu$1h${_{11/2}}{^{-1}}$1h${_{9/2}}{}$ multiplet shows gross mismatch with experiment for results with 
$sm56fp$ interaction. In the $sm56fph$, the corresponding $tbme$s relevant for 6${^+}_2$, 8${^+}_2$,  and 7${^+}_2$ states are tuned to reproduce them. Thus, with $sm56fph$ intearction, experimental energies of these states are reproduced much better (Fig. \ref{132sn}). 

The level at 4.352 MeV is  identified as a 3$^-$ state in the study by B. Fogelberg {\it et al.} \cite{Fogelberg_132sn}.
It is expected that the lowest $3^-_1$, $4^-_1$ and $5^-_1$ states originate from  the coupling of $\nu$2d${_{3/2}}{^{-1}}$2f${_{7/2}}{}$. Theoretically calculated energies of  $4^-_1$ and $5^-_1$ states match with experimental data within 100-200 keV for $sm56fph$ interaction,  without any tuning of the relevant $tbme$s, the $3^-_1$ state is over predicted by $\simeq 600 keV$. However,  $4^-_1$ originates from the partition 
$\nu$2d${_{3/2}}{^{-1}}$2f$_{7/2}$ (96\%), $5^-_1$ has 88\% contribution from $\nu$2d${_{3/2}}{^{-1}}$1h${_{9/2}}$. The $5^-_2$ calculated at 5.299 MeV is a member of the $\nu$2d${_{3/2}}{^{-1}}$2f${_{7/2}}$ multiplet. The inability of shell model theory to reproduce the energy of 
3$^-_1$ state  has been also discussed in Ref. \cite{Fogelberg_132sn}. They  have argued that the $3^-_1$ and $5^-_1$ levels of $^{132}$Sn should  be present within about 100 keV from each other.  The experimental energy of the $5^-_1$ level is 4.942 MeV. The experimental energy of  $3^-_1$ is around 600 keV lower than the other members of the $\nu$2d${_{3/2}}{^{-1}}$2f${_{7/2}}{}$ multiplet.  This is a signature of the collective nature of this state - which is generated from a coherent superposition of many p-h configurations. The failure of the present calculation (to be discussed in Sec. \ref{level3c}) to reproduce the enhanced E3 transition from this state also supports the possible collective nature of this state.

The $6^-_1$ state which originates from the partition $\nu$2d${_{3/2}}{^{-1}}$1h${_{9/2}}$ is also  underpredicted in theory.
However,  $6^-_2$, $6^-_3$ and $7^-_1$ states originated primarily from $\nu$2d${_{5/2}}{^{-1}}$2f${_{7/2}}$ (99\%), $\nu$2d${_{5/2}}{^{-1}}$1h${_{9/2}}$ (98\%) and $\nu$2d${_{5/2}}{^{-1}}$1h${_{9/2}}$ (95\%) partitions, respectively, are reproduced well in theory.
 
\begin{figure}
 \includegraphics[width =\linewidth]{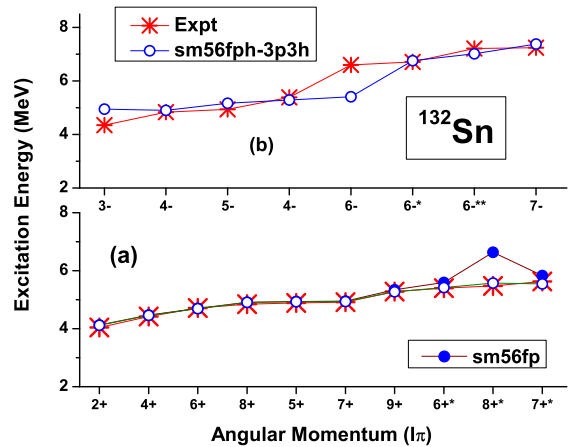}
 \caption{\label{132sn} (Color Online) Comparison of (a) positive parity and (b) negative parity states in the experimental spectra of $^{132}$Sn  with theory with 3p3h excitations. The results with $sm56fp$ are shown with filled blue bullets and those with $sm56fph$ are indicated with open blue bullets. The number of asterisks (N) appearing as superscript in the value of angular momentum in the x-axis label indicate it as the (N+1)th state of that spin.}
 \end{figure}

\subsubsection{\label{level3bb}$^{133}_{51}Sb_{82}$} 

The immediate neighbour of $^{132}$Sn, $^{133}$Sb with a proton coupled with the doubly closed shell nucleus,  provides excellent opportunity to study  the single proton states beyond Z=50. Its relatively higher spin, low excitation energy spectra  primarily involve neutron p-h excitation across the shell gap at N=82. Several experimental groups \cite{133sb_1973, 133sb_1978,133sb_urban_2000,133sb_1983_blmqvist,Genevey_2000} have worked  to extract experimental data for $^{133}$Sb. A high-spin isomer  arising from core-excitation was identified by them.   At GSI, using isochronous mass spectrometry, the core excited isomeric state was first directly identified at an excitation energy of  4.56(10) MeV with measured half-life of 17 $\mu$s 
\cite{133sb_sun_2010}. They have also provided a limiting value of the half-life for the fully ionized $^{133}$Sb nuclei corresponding to a totally disabled internal conversion decay mode. Information about the core-excited states of $^{133}$Sb above 21/2$^+$ state has been obtained later, from cold neutron induced fission of $^{241}$Pu and $^{235}$U in an experiment performed at ILL reactor in Grenoble \cite{133sb_bocchi_2016}.  The property of these core excited states, offer an opportunity to test the  newly formulated  shell model interaction.

The spectra of $^{133}$Sb are calculated with both 1p1h and 2p2h excitation options with $sm56fph$ interaction. No restriction has been put for neutrons 
in the N$<$ 82 orbitals. Results for both 1p1h as well as 2p2h are shown in the Fig. \ref{133sb}. The 3p3h excitations could not be included in the calculations due to computational limitation.

 In our calculation, 3/2$^+$ and 11/2$^-$ at the energy 2.44 MeV and 2.792 MeV are not  reproduced. These are originated from the single proton excitation to $\pi2d_{3/2}$ and $\pi1h_{11/2}$ orbitals, which are not included in the present model space. Thus these states are not included in the 
Fig. \ref{133sb}.

However, the higher energy ($>$ 4 MeV)  positive and negative parity  states arising from core-excitation are reproduced quite well (Fig. \ref{133sb})
in 2p2h excitation mode. Similar to  $^{132}$Sn, the  positive parity 11/2$^+_1$, 13/2$^+_1$, 15/2$^+_1$, 17/2$^+_1$ and 21/2$^+_1$ states 
originate  primarily  from the partition $\nu$1h${_{11/2}}{^{-1}}$2f${_{7/2}}{}$  coupled with $\pi$1g$_{7/2}$  with 90-94\% contribution in their wavefunctions. The 21/2$^+_2$, 23/2$^+_1$ and 25/2$^+_1$ have contribution (96\%, 89\% and 97\%) from the partition $\nu$1h$_{11/2}^{-1}$1h${_{9/2}}{}$  coupled with $\pi$1g$_{7/2}$.

The negative parity $13/2^-_1$ and $15/2^-_1$ states have 86\% and 70\% contributions from the partition  $\nu$2d$_{3/2}^{-1}$2f${_{7/2}}{}$ 
coupled with $\pi$1g$_{7/2}$.  The 2p2h option successfully reproduce experimental data except for $13/2^-_1$ and $21/2^+_2$ states, where 1p1h option is in better agreement.

\begin{figure}
\includegraphics[width =\linewidth]{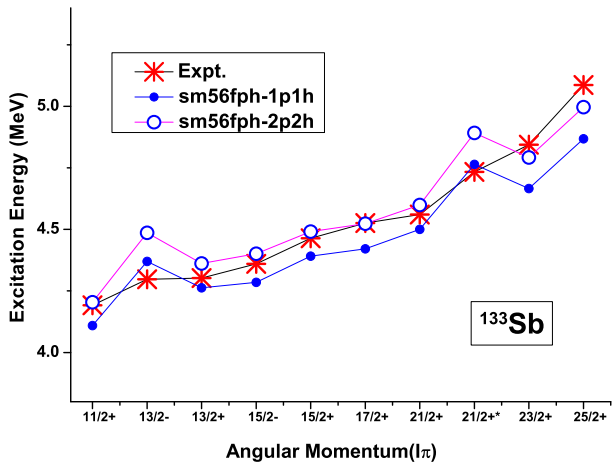}
\caption{\label{133sb}Comparison of experimental spectra  of high spin states in $^{133}$Sb with theory. More details are included in the legend, caption of Fig. \ref{132sn} and the 
text. }
\end{figure}
\subsubsection{\label{level3bc}$^{134}_{52}Te_{82}$} 

$^{134}$Te contains two valence protons outside the  $^{132}$Sn core. Most of the  experimental data  have been obtained from coincidence measurement of prompt and delayed gamma ray cascades of fission fragments emitted following spontaneous fission using large $\gamma$-detector arrays \cite{134te_fogelberg_1990,134te_omtvedt_1995,134te_135I_zhang_1996,134te_daly_1997,134te_135I_sksaha_2001}. 

At low excitations (upto $\simeq$ 3 MeV), the states  mainly originate   from proton excitations within the shell. The two most important partitions  are  $\pi 1g_{7/2}^2$ and $\pi 1g_{7/2}-2d_{5/2}$. Results from the present work with 2p2h excitations match quite well with experimental data even for these low spins (Fig. \ref{134te}). 
 
Above 3 MeV (but $<$ 4.5 MeV), few negative parity states are observed and these are expected to originate from the multiplets of $\pi$1g$_{7/2}$1h$_{11/2}$. These negative parity states were not reproduced. The present model space does not include $\pi 1 h_{11/2}$ orbit. Thus a neutron has to be excited beyond the shell gap, to produce a negative parity state,  which resulted in the  higher excitation energy.  These states are not included in the 
Fig. \ref{134te}.

\begin{figure}
\includegraphics[width =\linewidth]{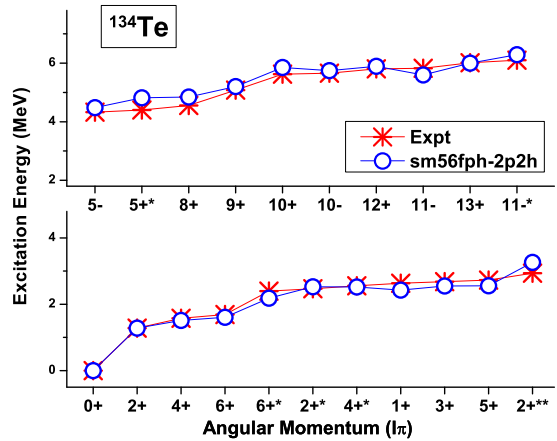}

\caption{\label{134te}Comparison of experimental spectra  of $^{134}$Te with theory. See legend, caption of Fig. \ref{132sn} and text for more details.}
\end{figure}

The first one in the set of excited states involving excitations across the neutron core has been identified at excitation energy 4.558 MeV \cite{134te_135I_sksaha_2001} with I$^\pi$=8$^+$. These positive parity states are interpreted to be members of the  multiplets of $\pi$1g$_{7/2}^2-\nu$1h$_{11/2}^{-1}$2f${_{7/2}}{}$. Results for 2p2h calculations are shown in the Fig. \ref{134te}. In our calculation, 8$^+_1$  state at 4.843 MeV  has this configuration with around 90\% amplitude. Whereas, the  9$^+_1$ has nearly 70\% contribution from $\pi$1g$_{7/2}^2-\nu$1h$_{11/2}^{-1}$1h${_{9/2}}{}$. The 10$^+_1$ state predicted at 5.85 MeV (E$_{expt}$= 5.621 MeV), has a mixed configuration with  $\pi$1g$_{7/2}^2-\nu$1h$_{11/2}^{-1}$2f${_{7/2}}{}$ ($\simeq$59\%), $\pi$1g$_{7/2}$2d$_{5/2}-\nu$1h$_{11/2}^{-1}$2f$_{7/2}$ ($\simeq$36\%). The yrast 12$^+$ (13$^+$) state  have  dominant contribution
of $\simeq$45\% (31\%) $\pi$1g$_{7/2}^2-\nu$1h$_{11/2}^{-1}$2f${_{7/2}}{}$  and $\simeq$48\% (65\%) $\pi$1g$_{7/2}$2d$_{5/2}-\nu$1h$_{11/2}^{-1}$2f${_{7/2}}{}$ partition. 

The negative parity state 10$^-_1$ (E$_{expt}$=5.658 MeV) at 5.746 MeV has  mixed composition with  45\% contribution from the partition
 $\pi$1g$_{7/2}^2-\nu$2d$_{3/2}^{-1}$ 2f$_{7/2}$ and 24\% from $\pi$1g$_{7/2}$2d $_{5/2}-\nu$2d$_{3/2}^{-1}$2f$_{7/2}$. The  11$^-_1$  has 77\% contribution from $\pi$1g $_{7/2}^2-\nu$2d$_{3/2}^{-1}$1h$_{9/2}$ and 11$^-_2$ on the other hand, has 72\% contribution from the partition 
 $\pi$1g$_{7/2}$2d$_{5/2}-\nu$2d$_{3/2}^{-1}$1h$_{9/2}$. Overall agreement with experimental data is reasonably  good.


\subsubsection{\label{level3bd}$^{135}_{53}I_{82}$}
The experimental data about the excited states of this nucleus have been mainly obtained  from  the decay \cite{135I_samri_1985}
of $^{135}$Te and from the  prompt $\gamma$ -ray spectroscopy of  the spontaneous fission fragments of $^{248}$Cm \cite{134te_135I_zhang_1996,134te_135I_sksaha_2001}. 

Three valence protons out side the Z=50 core dominate the low lying yrast excitations of this nucleus.  The low lying states within 
1.5 MeV (7/2$^+_1$, 9/2$^+_1$, 11/2$^+_1$, 15/2$^+_1$) have major contribution from  $\pi$1g$_{7/2}^3$. Theoretical results (Fig. \ref{135I}) show good agreement with data. The 5/2$^+_1$ and 17/2$^+_1$ states on the other hand show dominant participation of  $\pi$1g$_{7/2}{^{2}}$2d$_{5/2}$ partition. A few negative parity states are observed at  excitation energies beyond 3.5 MeV but less than 4 MeV. These are expected to arise from the multiplets of $\pi$1g$_{7/2}{^{2}}$1h$_{11/2}$, which is beyond the scope of this calculation.  Thus these states are not included in the Fig. \ref{135I}.

 The core excited states of $^{135}$I  have been observed after 4 MeV. Results with 1p1h excitations and 2p2h excitations  for $sm56fph$ 
interaction are shown in Fig. \ref{135I}. The  positive parity states (19/2$^+_1$, 21/2$^+_1$, and 23/2$^+_1$) are reproduced better with 2p2h excitations. The 19/2$^+_1$, 21/2$^+_1$ states have $\simeq$ 55\% and 42\% (57\% and 45\%) contributions from 
$\pi$1g$_{7/2}^2$2d$_{5/2}-\nu$1h${_{11/2}}{^{-1}}$2f$_{7/2}$ and $\simeq$ 36\% and 48\% (35\% and 46\%) from  
$\pi$1g$_{7/2}{^{3}} -\nu$1h${_{11/2}}{^{-1}}$2f$_{7/2}$ partitions  with  2p2h (1p1h) modes. 
For 23/2$^+_1$, both the modes of excitations predict $\simeq$ 75\% contribution from  $\pi$1g$_{7/2}{^{3}} -\nu$1h${_{11/2}}{^{-1}}$2f$_{7/2}$ partition.
Both the  modes predict 25/2$^+_1$ with a deviation of around 130 keV. They have common dominant structure of $\pi$1g$_{7/2}^3-\nu$1h${_{11/2}}{^{-1}}$1h$_{9/2}$ with $\simeq$ 83\% contributions. The Fig. \ref{135I} clearly shows the failure of $sn100pn$ interaction to reproduce the core-excited states.

\begin{figure}
\includegraphics[width =1.5\linewidth]{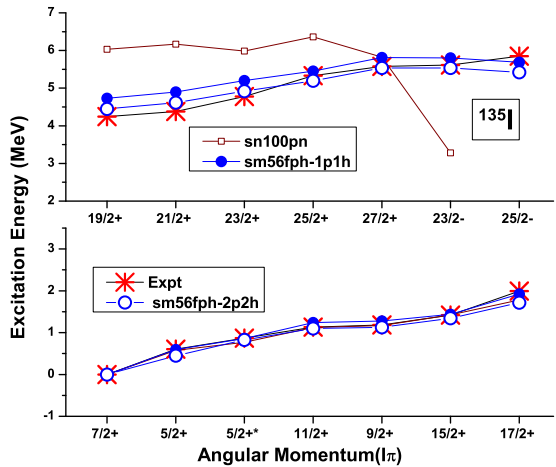}
\vspace{-3cm}
\caption{\label{135I}Comparison of experimental spectra  of $^{135}$I with theory. More details are included in the legend, caption of Fig. \ref{132sn} and the 
text.}
\end{figure}
Core excited negative parity states 23/2$^-_2$ (25/2$^-_1$) are observed after 5.7 MeV \cite{134te_135I_sksaha_2001}.
The negative parity state 23/2$^-_2$ shows better agreement with theoretical prediction for 2p2h excitation. The composition of the state has 
$\simeq$ 53\% (50\%) contribution from $\pi1g_{7/2}^2 2d_{5/2}-\nu 2d_{3/2}^{-1}2f_{7/2}$ for 1p1h (2p2h excitation). Experimentally observed 25/2$^-$ state at energy at 5.849 MeV matches with 25/2$^-_1$ state with $\simeq$ 67\% contribution from $\pi$1g$_{7/2}^2 2d_{5/2}$$-\nu$2d${_{3/2}}{^{-1}}$1h$_{9/2}$ for 1p1h excitation. Whereas, 2p2h shows deviation around 400keV with   $\simeq$ 57\% contribution from $\pi$1g$_{7/2}^2 2d_{5/2}$$-\nu$2d${_{3/2}}{^{-1}}$2f$_{7/2}$  partition.

\item {\it Calculations for N=81 isotones } 
\subsubsection{\label{level3be}$^{131}_{50}Sn_{81}$} 

This nucleus has a single neutron hole coupled to the doubly magic $^{132}$Sn. Thus  one can gain information about the single hole states, useful for  nuclear shell model calculations in its structure. Experimentally, the single hole states (1g$_{7/2}$, 2d$_{3/2}$, 3s$_{1/2}$) in $^{131}$Sn 
were studied from  $\beta$-decay experiments by De Geer and Holm \cite{131sn_geer_1980}. There is some uncertainty in  the relative position of 
the ${11/2^-_1}$ state originated from one neutron hole in 1h$_{11/2}$ orbit. This state is a $\beta$-decaying isomer. Several experiments have been performed to determine the energy of 11/2$^-_1$ state precisely \cite{131sn_fogelberg_1984_1,131sn_fogelberg_1984_2, 131sn_fogelberg_2004}. The most recently adopted data reports that the energy obtained  from the $\beta$ spectrum is 69 (14) keV, whereas it is 65.1 (3) keV from the level 
scheme \cite{nndc}. This state  is the first excited state of $^{131}$Sn nucleus. Our calculations for 3p3h mode of excitation with 
$sm56fph$ interaction predict this energy  as 36 keV. The trend of other low-lying excited states  1/2$^+_1$, 5/2$^+_1$, 7/2$^+_1$  at  0.332 MeV, 1.655 MeV, 2.434 MeV, respectively,  as determined from the decay studies \cite{131sn_geer_1980,131sn_fogelberg_2004} are reproduced in the calculated spectra
(Fig. \ref{131sn}). However, the theoretical values do not exactly reproduce the experimental data for 1/2$^+_1$ and 7/2$^+_1$.

Higher excited states in the energy range 4-5 MeV  populated from the spontaneous fission of $^{248}$Cm were studied via prompt-delayed gamma spectroscopy \cite{131sn_p_bhatta_132sn}.  The positive  parity states   13/2$^+_1$, 15/2$^+_1$, 17/2$^+_1$ and 19/2$^+_1$ are found to be originated from the coupling of   $\nu$2d${_{3/2}}{^{-1}}$1h${_{11/2}}{^{-1}}$ 2f$_{7/2}{}$ with the purity gradually increasing ranging from $\simeq$ 
{{82\% to 95\%}}. The second 15/2$^+_2$ state, although originates from the same partition has much less purity ($\simeq$ 61\%). The 17/2$^+_2$ 
evolves  from $\nu$2d${_{3/2}}{^{-1}}$1h${_{11/2}}{^{-1}}$1h$_{9/2}{}$ partition with 82\% contribution. 

The negative parity states 15/2$^-_1$, 17/2$^-_1$, 19/2$^-_1$, 21/2$^-_1$, and 23/2$^-_1$ have more pure structure with 94\%, 83\%, 96\%, 88\% and
94\% contributions, respectively,  from the $\nu$1h${_{11/2}}{^{-2}}$ 2f$_{7/2}{}$ partition.  These states are compared with theory by considering the excitation energy of the 11/2$^-_1$ isomer as zero.

\begin{figure}
\includegraphics[width =\linewidth]{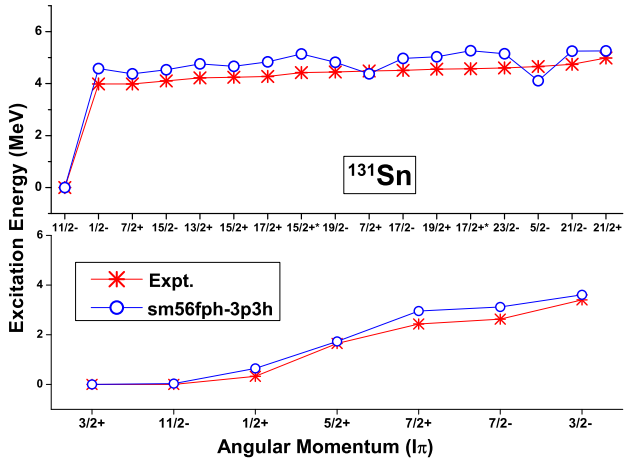}
\caption{\label{131sn}Comparison of experimental spectra  of $^{131}$Sn with theory. More details are included in the legend, caption of Fig. \ref{132sn} and the 
text.}
\end{figure}


\subsubsection{\label{level3bf}$^{132}_{51}Sb_{81}$}

This odd-odd isotope has one proton particle and one neutron hole with respect to the doubly closed $^{132}$Sn. At  low  excitation energies, the different states of this nucleus arise mainly from the interaction between the single proton and the single neutron hole. The  experimental data have been reported in Refs. \cite{132sb_Kerek_1972, 132sb_stone_1989, 132sb_mach_1995}.

The $4^+$ ground state and $3^+_1$ state have  been identified from earlier work to be composed of coupling between $\pi$1g$_{7/2}-\nu$2d$_{3/2}{^{-1}}$. The calculations reproduced the sequence as well the composition as per earlier reports (Fig. \ref{132sb}). The 3$^+_1$ state comes at an energy of {{135}} keV where the experimental energy is 85.55(6) keV. This state is also an isomer with a half-life of 15.62(13) ns.

The low lying $8^-_1$ isomer, whose experimental energy is yet to be determined has been produced at 193 keV in the calculations, with a dominant configuration as $\pi$1g$_{7/2}-\nu$1h$_{11/2}{^{-1}}$. This state is a beta decaying  isomer having a half-life of 4.10(5) min. 
   In the adopted levels \cite{nndc}, its  excitation energy is not specified but it was predicted to be around 150-250 keV from the ground state \cite{132sb_stone_1989}. In the atomic mass database \cite{data_audi}, this level is specified at 200 keV with an uncertainty of 30 keV.  The other members
of this  multiplet, J = $3^-_1$, $4^-_1$, $6^-_1$, $9^-_1$  are experimentally seen. In most of the cases, the purity of the structure is greater 
than {{95\%}}. 

 After 2.5 MeV, two positive parity states 10$^+$, 11$^+$ states were observed from the fission of $^{248}$Cm  by  P. Bhattacharyya {\it et al.} \cite{132sb_pb_2001}. These two states were identified to originate from coupling of  $\pi$1h$_{11/2}-\nu$1h$_{11/2}{^{-1}}$ with experimental energies  2.8 and 3.2 MeV, respectively. Because of absence of $\pi$1h$_{11/2}$ orbit, we could not reproduce these states. 

A few more positive   (11$^+$ ,12$^+$ ,13$^+$) as well as negative (12$^-$, 13$^-$, 14$^-$, 15$^-$) parity levels were identified at  excitation energies above 4 MeV. To generate spin more than 11, the core has to be excited i.e, one neutron has to be excited to the orbitals above the N=82 shell gap. These high spin states have been found to decay to the $8^-$ isomer, whose experimental energy is not known yet. Thus we have compared the experimental data with theoretical predictions considering the  $8^-_1$ (theo) state {{(i.e. 193 keV)}} to be zero. The 11$^+$  state is member of  the multiplet 
$\pi$1g$_{7/2}-\nu$1h$_{11/2}^{-1}$2d$_{3/2}^{-1}$2f$_{7/2}$ with 84\% parentage.  However, the other states 12$^+$, 13$^+$  are  coming from $\pi$1g$_{7/2}-\nu$1h$_{11/2}^{-1}$2d$_{3/2}^{-1}$1h$_{9/2}$ partition with 86\% and 90\% contributions. However, the core excited negative parity states 
(12$^-$ to 15$^-$) have around {{90-96\%}} parentage from  $\pi$1g$_{7/2}-\nu$1h${_{11/2}}{^{-2}}$2f$_{7/2}{}$ partition. 
 
 \begin{figure}
\includegraphics[width =1\linewidth]{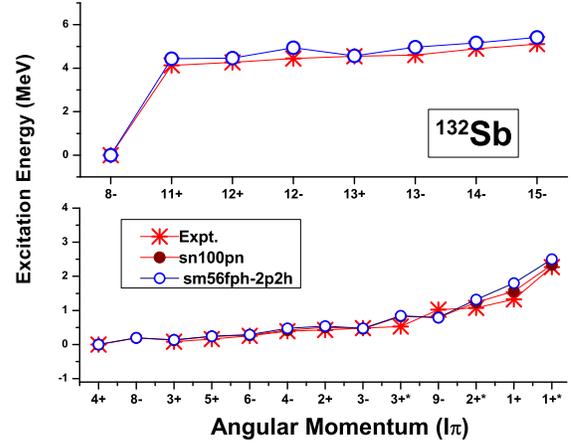}

\caption{\label{132sb}Comparison of experimental spectra  of $^{132}$Sb with theory. More details are included in the legend, caption of Fig. \ref{132sn} and the 
text.}
\end{figure}

 \begin{figure}
\includegraphics[width =1.\linewidth]{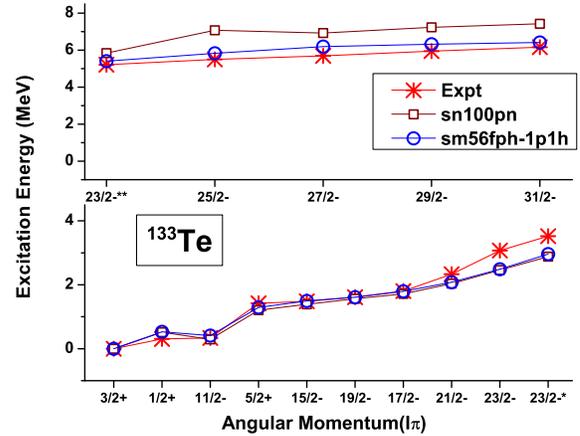}
\caption{\label{133te}Comparison of experimental spectra of $^{133}$Te with theory. More details are included in the legend, caption of Fig. \ref{132sn} and the 
text.}
\end{figure}

\subsubsection{\label{level3bg}$^{133}_{52}Te_{81}$}

$^{133}$Te can be described as $^{132}$Sn with additional two protons and one neutron hole. The second excited state (11/2$^-$) at 0.334 MeV is a beta 
decaying isomer with T$_{1/2}$ = 55.4 (4) min.  It is theoretically predicted at 0.413 MeV. The experimental details of the $3/2^+$ ground state, the 
1/2$^+$ state below this isomer  and other low-spin states have been found from $\beta^-$ decay of $^{133}$Sb \cite{nndc}. 

The theoretical results are compared with experimental data in Fig. \ref{133te}. The positive parity states 3/2$^+_1$ and 5/2$^+_1$ have contributions  
93\% and 74\%, respectively, from the partition  $\pi$1g$_{7/2}^2-\nu$2d$_{3/2}^{-1}$. On the other hand, 1/2$^+$ state is originated from the partition $\pi$1g$_{7/2}^2-\nu$3s$_{1/2}^{-1}$ with 74\% amplitude.  
 
 The yrast isomer 11/2$^-$ and other members of the sequence 15/2$^-_1$, 17/2$^-_1$, 19/2$^-_1$,  23/2$^-_1$ originate from  $\pi$g$_{7/2}^2$ coupled with $\nu$h$_{11/2}^{-1}$ with more than 90\% contribution.  The yrast 21/2$^-$ has a mixed structure with 52\% contribution from the partition 
$\pi$g$_{7/2}^2$-$\nu$h$_{11/2}^{-1}$ and 46\% from $\pi$1g$_{7/2}$2d$_{5/2}-\nu$1h$_{11/2}^{-1}$. The 23/2$^-_2$ state originates from 
$\pi$1g$_{7/2}$2d$_{5/2}-\nu$1h$_{11/2}^{-1}$ (98\%) as also suggested by Hwang {\it et al.} \cite{hwang_2002}.

The positive states near 4 MeV, (21/2$^+$, 25/2$^+$, 23/2$^+$, 27/2$^+$) are not reproduced well in the present calculations. Hwang {\it et al.} 
\cite{hwang_2002} have identified these states to originate from  $\pi$1g$_{7/2}$1h$_{11/2}-\nu$1h$_{11/2}^{-1}$. Thus these  energies could not be reproduced in the present calculations  because of the absence of the proton 1h$_{11/2}$ orbit in the model space.

Hwang et al.\cite{hwang_2002} interpreted the sequence of states  beginning at 5.2 MeV in $^{133}$Te as originated from the particle-hole excitation
from the $^{132}$Sn core. From our calculation, the  23/2$^-_3$ and 25/2$^-_1$ have major contribution (86\% and 71\%, respectively) from the partition $\pi$1g$_{7/2}{^2}-\nu$1h$_{11/2}^{-2}$2f$_{7/2}^{-1}$. However, the other band members, {\it viz.},  27/2$^-_1$, 29/2$^-_1$ and 31/2$^-_1$ have contributions from both $\pi$1g$_{7/2}$2d$_{5/2}-\nu$1h$_{11/2}^{-2}$2f$_{7/2}^{-1}$  (61-75\%) and  
$\pi$1g$_{7/2}{^2}-\nu$1h$_{11/2}^{-2}$2f$_{7/2}^{-1}$ (30-21\%). As expected, the results with $sn100pn$ interaction show substantial deviation from experimental data.

\item {\it Calculations for N=83 isotones } 

\subsubsection{\label{level3bh}$^{134}_{51}Sb_{83}$}
 \begin{figure}
\includegraphics[width =1.5\linewidth]{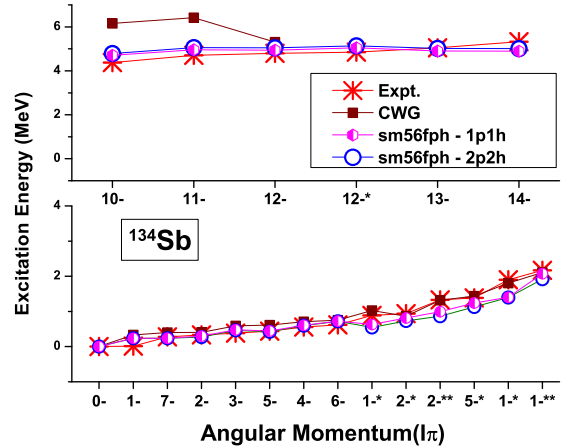}
\vspace{-4cm}
\caption{\label{134sb}Comparison of experimental spectra of $^{134}$Sb with theory. More details are included in the legend, caption of Fig. \ref{132sn} and the 
text.}
\end{figure}

$^{134}$Sb can be characterised as one proton and one neutron particle outside the $^{132}$Sn core. The Fig. \ref{134sb} shows the comparison of theoretical results (done with $CWG$ interaction and 1p1h and 2p2h excitations modes with $sm56fph$ interaction) with experimental data. Both the modes reproduce the experimental data for higher spin states much better than those predicted by CWG.

$^{134}$Sb  has a beta decaying isomer with spin  7$^-$ having T$_{1/2}$ = 10.07 (5)s. The details of low lying structure have been obtained from  
${\beta}^-$ and ${\beta}^-$n decay of $^{134,135}$Sn, respectively \cite{shergur_2002}. There is an experimentally observed  1$^-$ state almost degenerate to 0$^-$ at 13 keV. Both $CWG$ and $sm56fph$ interactions  are unable to reproduce the low-lying 1$^-$ state. The yrast 0$^-$ to 7$^-$ states have dominant contribution of  the partition $\pi$1g$_{7/2}-\nu$2f$_{7/2}$. The  second 1$^-$ state originate from the partition $\pi$1g$_{7/2}-\nu$1h$_{9/2}$. The 2$^-_2$ and 2$^-_3$ have around 70\% contributions from  $\pi$1g$_{7/2}$ coupled with neutron in 1h$_{9/2}$ and 3p$_{3/2}$, respectively. The 5$^-_2$ state has a leading contribution 59\% from the partition  $\pi$1g$_{7/2}-\nu$3p$_{3/2}$. Near 2 MeV, 1$^-_3$ and 1$^-_4$ have  proton in  2d$_{5/2}$ coupled with neutron in 2f$_{7/2}$ (94\%) and 3p$_{3/2}$ (95\%), respectively. 

After 2 MeV, the positive states experimentally observed are 9$^+$ and 10$^+$ states which have dominant contribution from the partitions
 $\pi$1h$_{11/2}-\nu$2f$_{7/2}$  and $\pi$1g$_{7/2}-\nu$1i$_{13/2}$, respectively, as discussed in Ref. \cite{Fornal_2001}. But, our calculation predict these states at higher energies because of unavailability of proton 1h$_{11/2}$  and neutron 1i$_{13/2}$ orbits in the model space. 

The high spin negative parity states arise from the neutron core breaking. In our calculation in 2p2h mode, the yrast 10$^-$ state has been reproduced
with leading contribution (73\%) from the partition  $\pi$1g$_{7/2}-\nu$1h$_{11/2}^{-1}$2f$_{7/2}^2$. The other negative parity  yrast 
11$^-$ to 14$^-$ states have gradually increasing  contribution  of the partition $\pi$1g$_{7/2}-\nu$1h$_{11/2}^{-1}$2f$_{7/2}$1h$_{9/2}$ ranging from 49\% to 92\%. This structure is different from that ($\pi$1g$_{7/2}-\nu$1h$_{11/2}^{-1}$2f$_{7/2}^2$) suggested by 
Fornal {\it et al.} in Ref \cite{Fornal_2001}.


\subsubsection{\label{level3bi}$^{135}_{52}Te_{83}$}
$^{135}$Te  consists of two protons and one neutron outside  doubly magic $^{132}$Sn. The low lying states of this nucleus were studied mainly from the beta decay studies by Hoff et al\cite{hoff_1989}. The information about the isomeric state is known from Kawade et al.\cite{kawade_1980}. The investigation regarding the high spin states are known from Refs. \cite{bhatta_1997,Fornal_2001,luo_2001}. The Fig. \ref{135Te} shows the comparison of theoretical results (done with $CWG$ interaction and 1p1h  excitations mode with $sm56fp$ and $sm56fph$ interactions) with experimental data. Both the modes reproduce the experimental data for higher spin states much better than those predicted by CWG.

For $sm56fp$, low lying states (7/2$^-$, 11/2$^-$, 15/2$^-$) are originated from the partition $\pi$1g$_{7/2}^2- \nu$2f$_{7/2}$ with contributions approximately 94, 91 and 82\%, respectively. Same is observed for $sm56fph$ with  contributions approximately  93, 87 and 82\%, respectively. The yrast
3/2$^-$ state has 75\% contribution of the partition $\pi$ 1g$_{7/2}^2 -\nu$3p$_{3/2}$ for $sm56fp$ with  and 73\% for $sm56fph$. For 1/2$^-$ and 
5/2$^-$ states, the dominant partition is  $\pi$ 1g$_{7/2}^2 -\nu$2f$_{7/2}$ and it contributes 78\% (77\%) and 71\% (65\%), respectively, for 
$sm56fp$ ($sm56fph$) interaction. For $sm56fph$ interaction,  the 9/2 $_1^-$ state has 95\% contribution from the partition 
$\pi$ 1g$_{7/2}^2 -\nu$2f$_{7/2}$. However, this yrast state has 97\% contribution from  $\pi$ 1g$_{7/2}^2- \nu$2f$_{7/2}$ for $sm56fp$ interaction.

Yrast  19/2$^-$ state originates from the partitions $\pi$ 1g$_{7/2}^2 -\nu$2f$_{7/2}$ with 47\%(48\%)  and  $\pi$ 1g$_{7/2}$2d$_{5/2} -\nu$2f$_{7/2}$ with 53\%(50\%) contributions, for $sm56fp$ ($sm56fph$) interaction. The second 19/2$^-$ has  $\pi$ 1g$_{7/2}^2 -\nu$2f$_{7/2}$ partition
with 53\%(50\%)  and  $\pi$ 1g$_{7/2}$2d$_{5/2} -\nu$2f$_{7/2}$ with 47\%(48\%) contribution for $sm56fp$ ($sm56fph$) interaction. The 
19/2$^-_3$ has dominant contribution of 78\% from the partition $\pi$1g$_{7/2}^2 -\nu$1h$_{11/2}^{-1}$2f$_{7/2}{^2}$ for $sm56fp$. On the other hand, in case of $sm56fph$ it has a contribution of 98\% from $\pi$1g$_{7/2}^2 -\nu$1h$_{9/2}$ partition. 

The yrast 21/2$^-$ has 52\% contribution from  $\pi$1g$_{7/2}^2$2d$_{5/2}- \nu$1h$_{11/2}^{-1}$2f$_{7/2}{^2}$ partition, in case of $sm56fp$ interaction.  But for $sm56fph$ interaction the dominant contribution is coming from $\pi$1g$_{7/2}^2 -\nu$1h$_{9/2}$ (86\%).  The 21/2$^-_2$ state has 
$\pi$1g$_{7/2}^2 -\nu$1h$_{11/2}^{-1}$2f$_{7/2}{^2}$ with  52\% along with  competing contribution of 36\% from $\pi$1g$_{7/2}$2d$_{5/2} -\nu$1h$_{11/2}^{-1}$2f$_{7/2}{^2}$ for $sm56fp$. For $sm56fph$,  this state mainly arises from $\pi$1g$_{7/2}$2d$_{5/2} -\nu$1h$_{9/2}$ with 86\% contribution.  

The 23/2$^-_1$, 25/2$^-_1$, 27/2$^-_1$, 29/2$^-_1$, 31/2$^-_1$, 33/2$^-_1$ and 35/2$^-_1$ states have contributions from  $\pi$1g$_{7/2}^2 -\nu$1h$_{11/2}^{-1}$2f$_{7/2}{^2}$ and $\pi$1g$_{7/2}$2d$_{5/2} -\nu$1h$_{11/2}^{-1}$2f$_{7/2}{^2}$ with amplitudes ranging from 41\% - 13\% and 31\% - 86\% for 
$sm56fp$ interaction.        \\

For $sm56fph$ interaction, 23/2$^-_1$, 25/2$^-_1$, 27/2$^-_1$, 29/2$^-_1$, 31/2$^-_1$, 33/2$^-_1$ and 35/2$^-_1$ have a mixed structure of partitions $\pi$1g$_{7/2}^2-\nu$1h$_{11/2}^{-1}$2f$_{7/2}{}$1h$_{9/2}{}$ and $\pi$1g$_{7/2}$2d$_{5/2}-\nu$1h$_{11/2}^{-1}$2f$_{7/2}{}$1h$_{9/2}{}$ with amplitudes ranging from 86\%-25\% and 2-74\%. 

Some of the positive parity states (say, 21/2$^+$, 25/2$^+$, 27/2$^+$ etc.) could not be reproduced well with these two interactions because of absence of proton h$_{11/2}$ orbital in the model space. 
\end{itemize}

 \begin{figure}
\includegraphics[width =1.\linewidth]{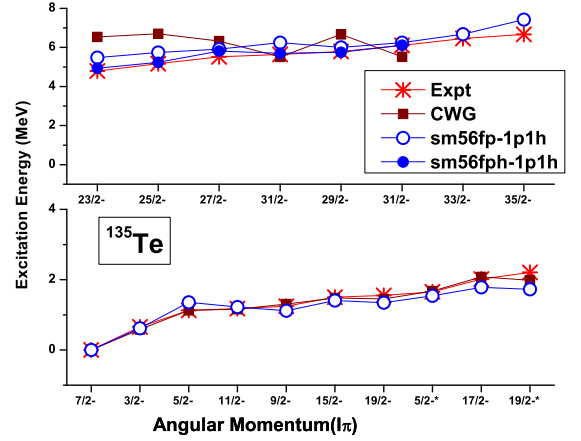}

\caption{\label{135Te}Comparison of experimental spectra of $^{135}$Te with theory. More details are included in the legend, caption of Fig. \ref{132sn} and the 
text.}
\end{figure}

\subsection{\label{level3c}Transition Probabilities in $^{132}$Sn and its neighbours
}
The transition probabilities of  yrast states which arise from cross-shell excitations in $^{132}$Sn, $^{131}$Sn, $^{133}$Sb and $^{134}$Te isotopes, for which experimental data \cite{nndc} are available, have been calculated (Table \ref{ball}) for $sm56fph$  and compared with experimental data 
\cite{nndc}. In case of E$\lambda$ ($\lambda$=2-4) the neutron effective charge is taken  as 1.2$e$. The results for all modes of excitation, which were calculated are shown in the table. However, the results for different modes are almost similar. 

The  results reproduce data reasonably well for $^{132}$Sn, except in  a few cases (8$^+$ to 6$^+$ and those involving the 3$^-$ state. For magnetic transitions, quenched intrinsic neutron g-factor  ($g_n^s$) by a factor of 0.7, improves the results.

Gross mismatch is observed for the E3 transition rate from the $3^-_1$ state. It has been already discussed that this state  possibly has a collective structure and thus its energy as well as transition rates are not reproduced in this calculation with such limited options for excitations.

In case of E1 transition, effective charge has been taken as e${_n}^{eff}$= 0.35$e$ and e${_p}^{eff}$= 1.64$e$ (Table \ref{ball}).  However,  the calculated values are underestimated compared to experimental data by   one order of magnitude in most cases.  However for E1 transitions from $5^-_1 \rightarrow 4^+$ state, the B(E1) values are well reproduced. The B(E1) value for $5^-_1 \rightarrow 6^+$ state  state is overpredicted compared to experimental data. 

The 2f$_{5/2}$,   3p$_{1/2}$, and 1i$_{13/2}$ orbits have been excluded from the model space. They are not usually significant for reproduction of energy eigenvalues or transition probabilities of the nuclei which are being discussed in this work. 
However this argument does not hold for E1 transition probabilities, {\it i.e. } for B(E1) values. These transitions  occur due to (very) small 
contributions of the orbitals originating in higher shells in  the wavefunction. Thus small overlap in wave-functions are also significant for proper reproduction of experimental B(E1)s, which are  always at least 5-6 orders of magnitude slower than the single particle estimate. 
 
For $^{131}$Sn, the B(E2) value for the  23/2$^-$ to  19/2$^-$ transition has been reproduced with the same effective charge to a reasonable extent.
However, the B(E1) and B(E4) values are under-predicted. 

For  $^{133}$Sb, the sharp difference between experimental B(M1) values extracted for the $15/2^+ \rightarrow  13/2^+$ and $13/2^+ \rightarrow 11/2^+$ transitions could not be reproduced for the calculated yrast states for both 1p1h and 2p2h modes. However, 2p2h mode reproduces their energies quite well. This is similar to the observation of Wang {\it et al.} \cite{wang}.  However, the calculated   B(E2) value of the $21/2^+ \rightarrow  17/2^+$ transition from the microsecond 21/2$^+$  isomer (without the internal conversion correction) also yields a milli-second half life of the isomer as  reported by B. Sun {\it et al.} \cite{133sb_sun_2010}.  

In $^{134}$Te, calculated B(E2) for the core excited 12$^+$ is under-predicted in theory.


\begin{table*}
\begin{center}
\caption{\label{ball}The comparison of theoretical transition rates of yrast states which arise from cross-shell excitations 
in $^{132}$Sn, $^{131}$Sn, $^{133}$Sb and $^{134}$Te isotope with experimental data \cite{nndc}.   
The  B(M1)  values are quoted in $\mu{_n{^2}}$. The  B(E$\lambda$)  values are quoted in unit of $e^2fm^{2\lambda}$. 
 The B(E1) values are quoted in the unit of $10^{-6} e^2fm^2$.} 

\begin{tabular}{ccccccc c c cc}
\hline\\
Nucleus &J$_i$ & J$_f$ &E$_\gamma$        & Type &Expt.     &\multispan{5} \hfil Theory\hfil \\
        &      &        &(keV)       &      &      (Error)    &\multispan{2}Effective charges &\multispan{3}\hfil Mode  \hfil  \\
        &      &        &       &      &          &$e_p$&$e_n$&1p1h  &2p2h&3p3h  \\

\hline
\\
$^{132}$Sn\\
&2$^+$ & 0$^+$     &4041.1 & E2       & 224.6 (6)     & -& 1.2& 235.3  & 232.8 &232.9  \\
&4$^+$ & 2$^+$     & 375.1& E2       &16.3 (10)      & -& " & 3.4&3.3& 3.0 \\
&6$^+$ & 4$^+$     & 299.6 &  E2         & 11.9 (4)   & -& "& 8.1  & 8.3 & 8.0  \\
&8$^+$ & 6$^+$     & 132.5 & E2        & 4.2 (1)      & -& "& 0.49  &0.48 &0.26 \\
&5$^-$ & 3$^-$     & 590.6  & E2          & 24.9(3)      & -& "& 0.09 & 0.041& 0.05 \\
\\
&3$^-$ & 0$^+$     &4351.9 & E3           & $>$7348.4    & -& 1.2& 82.3 & 81.5& 71.5  \\
\\
&4$^+$ & 0$^+$     & 4416.2 & E4        & 227.1(4)$\times 10^3$  & -& 1.2& 153 $\times$ 10$^3$ & 154 $\times$ 10$^3$  &149 $\times$ 10$^3$  \\
\\
&3$^-$ & 2$^+$   & 310.7 & E1& $>$ 284  & 1.64& 0.35& - & 0.31& 0.83   \\
&4$^+$ & 3$^-$  & 64.4 &E1       &  4.45(55)&  "&  "& -& 0.055 & 0.12\\
&4$^-$ & 4$^+$  & 414.6 & E1       &  4.85(48)&  "&  "&-& 0.29 &0.23 \\
&5$^-$ & 6$^+$  & 226.7 &E1       &  4.90(53)    &  "&  "&-&25.3& 26.4   \\
&5$^-$ & 4$^+$   & 526.2 &E1      &  140(15)   &  "&  "&-& 113.8& 120.1\\
\\
&7$^+$ & 8$^+$     & 70.4 &  M1       & 0.042(5)  &\multispan{2}\hfil 0.7g$_n{^s}$\hfil&0.019& 0.034& 0.033  \\
&7$^+$ & 6$^+$     & 203.1 & M1        & 0.066(7) &\multispan{2}\hfil 0.7g$_n{^s}$\hfil&0.029& 0.046& 0.045 \\
&5$^-$ & 4$^-$     & 111.5 & M1        & 0.123(14) &\multispan{2}\hfil 0.7g$_n{^s}$\hfil&0.028& 0.023& 0.020 \\

\hline\\
$^{131}$Sn \\
         & 19/2$^-$  & 17/2$^+$& 173.185 &E1& $>$ 333{\bf }&1.64 &0.35& - &0.13&0.40\\\\
         & 23/2$^-$   & 19/2$^-$ & 158.50 &E2 & 14.7 (10)          &  -      &1.2& 9.43  & 9.60&9.1 \\\\
         &19/2$^-$     & 11/2$^-$ &4446.0&E4 &$>$ 1.891$\times 10^{6}$  &  -  &1.2&  1.069$\times 10^{5}$ &1.078$\times 10^{5}$& 1.03$\times 10^{5}$     \\
            \hline\\
$^{133}$Sb\\
& 21/2$^+$ & 17/2$^+$ &$<$ 20&E2 &  $\sim$ 10.48    &  1.85      &0.35& 1.16  &1.17 - \\\\

&15/2$^+$ & 13/2$^+$& 162.3&M1   &$>$  0.43  &\multispan{2}\hfil free g${^s}$\hfil & 0.19 & 0.17& - \\
&13/2$^+$ & 11/2$^+$ & 110.2&M1  &  0.0075 (27) &\multispan{2}\hfil" \hfil  & 0.58 & 0.61& -  \\
\hline\\
$^{134}$Te\\ & 12$^+$   & 10$^+$& 182.6 &E2     & 133   (12)& 2.0&0.0   &36.8 &41.1& -\\
\hline\\
%

\end{tabular}
\end{center}
\end{table*}

\section{\label{sec:level4}Summary and Conclusion}
The excitation spectra of $^{132}$Sn and its nearest neighbours  have  been reproduced with a new cross-shell interaction. A few two body matrix elements have been tuned to reproduce the low-lying multiplet states of  $^{132}$Sn. This is the first full scale shell model study of $^{132}$Sn energy spectra as well as transition probabilities. The excitation spectra for other nearest neighbours are reproduced well. The transition probabilities of core-excited states are also reproduced reasonably well.  The most important observable calculated are the E1 transition probabilities for nuclei below and above $^{132}$Sn.  Interestingly,  the B(E1) rates from $5^-_1$ state in $^{132}$Sn are reproduced reasonably well. However, other  calculated  B(E1) values are systematically underpredicted. It indicates the need for inclusion of other higher energy neutron orbits in the model space. The new interaction will have far reaching consequences in explaining low -lying E1 isomers in mid-shell Sn and other nuclei  with N$<82$ as well  in those beyond N=82.

\end{document}